\documentclass[useAMS,usenatbib]{mn2e}

\usepackage{graphicx}
\usepackage{epsfig}
\newcommand{\LA}{\mbox{\raisebox{-0.6ex}{$\stackrel{\textstyle<}{\sim}$}}}

\begin{document}
\title{The Extreme Ultraviolet Spectra of Low Redshift Radio Loud Quasars}
\author[Brian Punsly Cormac Reynolds Paola Marziani and Christopher P. O'Dea] {Brian Punsly  Cormac Reynolds Paola Marziani and Christopher P. O'Dea\\ 1415 Granvia Altamira, Palos Verdes Estates CA, USA
90274 and ICRANet, Piazza della Repubblica 10 Pescara 65100, Italy\\
\\ Australia Telescope National Facility, CSIRO Astronomy and Space
Science, 26 Dick Perry Ave., Kensington, WA 6151, Australia\\
\\INAF, Osservatorio Astronomico di Padova, Italia\\
\\Department of Physics and Astronomy, University of
Manitoba, Winnipeg, MB R3T 2N2 Canada and School of Physics\\
\\ and Department of Astronomy, Rochester Institute of Technology, Rochester,
NY 14623, USA\\
 \\E-mail: brian.punsly1@verizon.net}

\maketitle \label{firstpage}
\begin{abstract} This paper reports on the extreme
ultraviolet (EUV) spectrum of three low redshift ($z \sim 0.6$)
radio loud quasars, 3C 95, 3C 57 and PKS 0405-123. The spectra were
obtained with the Cosmic Origins Spectrograph (COS) of the Hubble
Space Telescope. The bolometric thermal emission, $L_{bol}$,
associated with the accretion flow is a large fraction of the
Eddington limit for all of these sources.  We estimate the long term
time averaged jet power, $\overline{Q}$, for the three sources.
$\overline{Q}/L_{bol}$, is shown to lie along the correlation of
$\overline{Q}/L_{bol}$ and $\alpha_{EUV}$ found in previous studies
of the EUV continuum of intermediate and high redshift quasars,
where the EUV continuum flux density between 1100 \AA\, and 700
\AA\, is defined by $F_{\nu} \sim \nu^{-\alpha_{EUV}}$. The high
Eddington ratios of the three quasars extends the analysis into a
wider parameter space. Selecting quasars with high Eddington ratios
has accentuated the statistical significance of the partial
correlation analysis of the data. Namely. the correlation of
$\overline{Q}/L_{\mathrm{bol}}$ and $\alpha_{EUV}$ is fundamental
and the correlation of $\overline{Q}$ and $\alpha_{EUV}$ is spurious
at a very high statistical significance level (99.8\%). This
supports the regulating role of ram pressure of the accretion flow
in magnetically arrested accretion models of jet production. In the
process of this study, we use multi-frequency and multi-resolution
Very Large Array radio observations to determine that one of the
bipolar jets in 3C 57 is likely frustrated by galactic gas that
keeps the jet from propagating outside the host galaxy.
\end{abstract}
\section{Introduction}
There is mounting evidence that radio jet power is anti-correlated
with the extreme ultraviolet (EUV) luminosity in quasars
\citep{pun14,pun15}. In particular, the three quantities, the long
term time averaged jet power, $\overline{Q}$, the bolometric thermal
emission, $L_{bol}$, associated with the accretion flow and
$\alpha_{EUV}$, the spectral index of the EUV continuum (the EUV
continuum flux density is defined as the flux density between 1100
\AA\, and 700 \AA\, where $F_{\nu} \sim \nu^{-\alpha_{EUV}}$) are
inter-related. A partial correlation analysis indicted that the
fundamental correlation is between $\overline{Q}/L_{bol}$ and
$\alpha_{EUV}$ \citep{pun15}. Furthermore, evidence of a real time
anti-correlation between the contemporaneous jet power, $Q(t)$, and
the EUV luminosity was shown over a $\sim$ 40 year period for the
distant quasar, 1442+101 \citep{pun16}.

\par All of these results were developed based on intermediate and
high redshift quasars ($z > 0.65$) due to the constraint that the
EUV continuum at 700\AA\, is observable with the Hubble Space
Telescope (HST) for $z > 0.65$. We believe that by studying the EUV
spectra of very bright quasars in the EUV, we can push this limit a
little further to slightly lower redshift. In this paper, we
consider three low redshift radio sources at $z\sim 0.6$, 3C 57, 3C
95 and PKS 0405-123 \footnote{With the Hubble Space Telescope, the
shortest wavelength that can be efficiently detected with the Cosmic
Origins Spectrograph is $\lambda_{o} \approx 1150 \AA\ $. This
corresponds to an emitted wavelength of $\lambda_{e} \approx
1150/(1+z) \AA\ $. For $z = 0.6$, $\lambda_{e} \approx 719 \AA\ $.
Typically, in quasar spectra, there is a strong very broad NIV,
NeVIII emission line centered at $\sim 765\AA\, -780\AA\,$ which
will obfuscate the continuum unless the blue wing is completely
resolved \citep{shu12,pun15}. Thus, $z \LA 0.6$ is the extreme limit
of our ability to reliably estimate $\alpha_{EUV}$ in the range
$1100 \AA\,
> \lambda_{e} >700 \AA\ $.}. These quasars have been selected to be
observed in the EUV because they are the most EUV luminous radio
loud quasars (RLQs) in the local Universe. This allows us to probe
the $\overline{Q}/L_{bol}$ and $\alpha_{EUV}$ relationship at low
redshift and high Eddington rate. Fortunately, there are archival
high sensitivity Cosmic Origins Spectrograph observations of these
three sources. The high EUV luminosity and the high quality spectra
allow us to reliably estimate $\alpha_{EUV}$ even though the
spectral data ends at $\approx 730\AA\,$.
\par In next section, we present the EUV spectra of the three
sources.  In Section 4, we derive $\overline{Q}$ for these quasars
and in Section 5, the results are discussed in terms of the
$\overline{Q}/L_{bol}$ and $\alpha_{EUV}$ correlation.
\begin{figure*}
\includegraphics[width=110 mm, angle= 0]{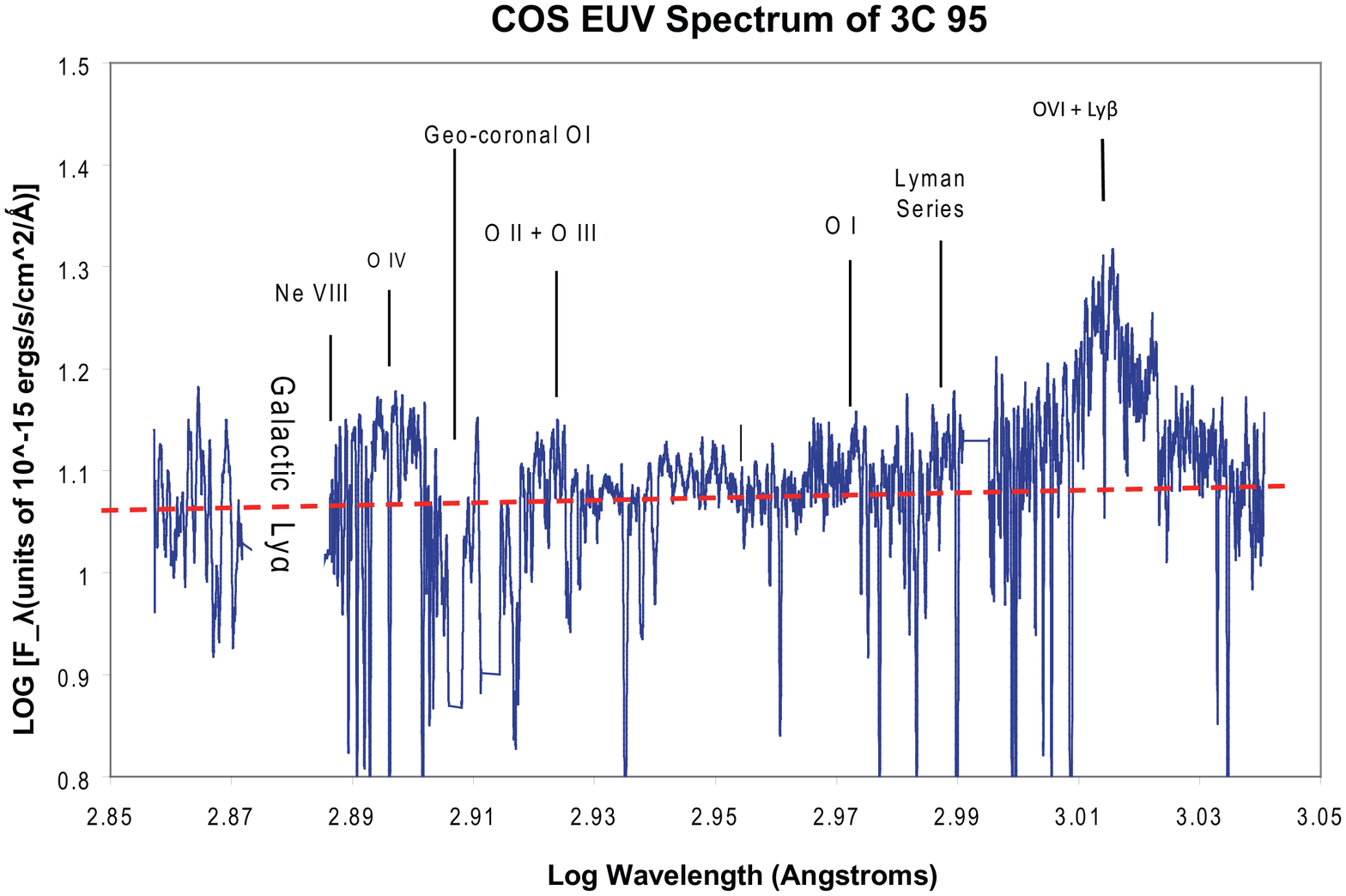}
\includegraphics[width=110 mm, angle= 0]{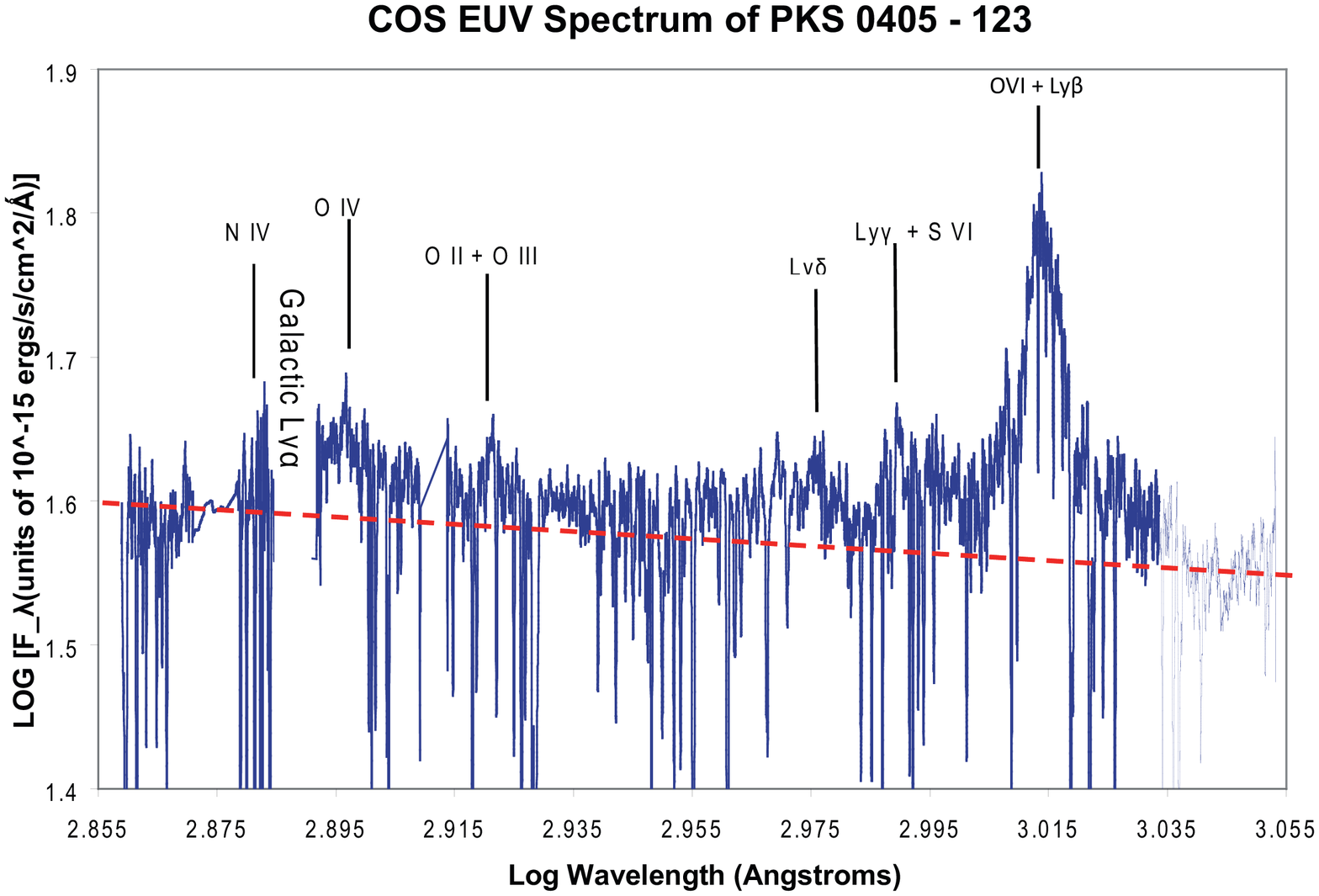}
\includegraphics[width=110 mm, angle= 0]{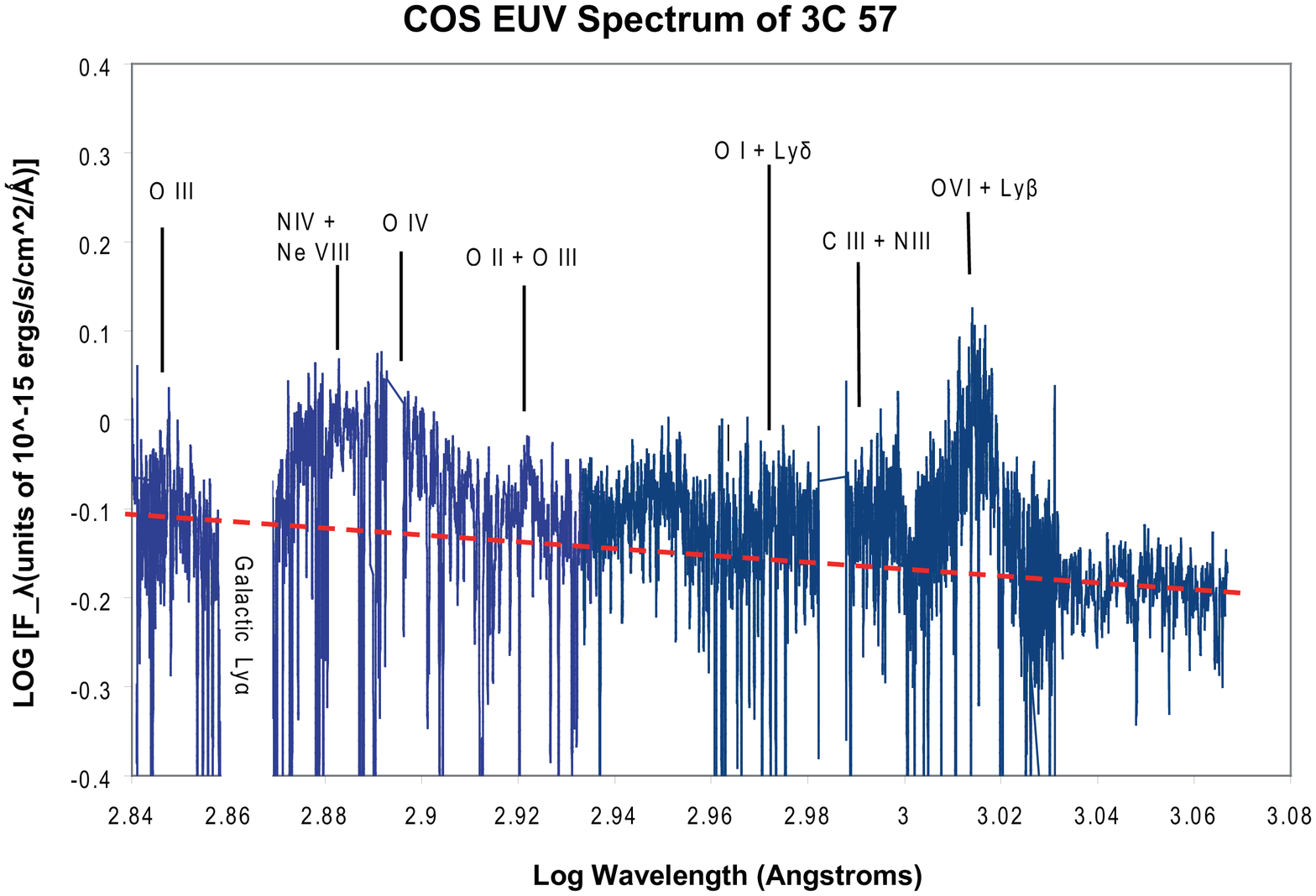}
\caption{The EUV spectra of 3C 95 (top), PKS 0405-123 (middle) and
3C 57 (bottom). the red dashed line is the fit to the EUV
continuum.}
\end{figure*}

\section{The EUV Spectra} In this section, the EUV spectra of the
three RLQs are described. All spectra are from the COS spectrograph
on HST with the G130M and G160M gratings. The flux density data that
are downloaded from MAST are in units of ergs/sec/$\rm{cm}^{2}$/\AA;
$F_{\lambda}$. The spectral indices of the EUV continuum derived
from composite spectra in \citet{tel02,zhe97} were based on
$F_{\nu}$. Consequently, the original studies on the EUV continuum
of radio loud quasars, \citet{pun14,pun15,pun16} adopted the same
convention and we continue this here for the sake of easy comparison
to previous treatments. Hence, our definition of the continuum power
law, $F_{\nu} \sim \nu^{-\alpha_{EUV}}$. The MAST data will gives us
the power law fit as $F_{\lambda} \sim \lambda^{\alpha_{\lambda}}$.
We convert this to the desired form with the formula, $\alpha_{EUV}=
2 + \alpha_{\lambda} $.
\subsection{3C 95} The quasar 3C 95 is a classic relaxed lobe dominated
quasar at a redshift, z = 0.616 that extends more than 500 kpc from
end to end \citep{hut88,pri93}. The H$\beta$ profile is very wide
with a full width half maximum of the broad component equal to 9030
\AA\, (see Table 1). The details required for an estimate of the
central black hole mass are summarized in Table 1. Columns (2) and
(3) are the data needed for an estimate of the black hole mass in
column (6) using the mass estimator in \citet{ves06} and the very
broad H$\beta$ profile. Columns (4) and (5) provide the data needed
for the independent virial mass estimator of \citet{she12} based on
the MgII profile. This result is listed in the last column. The data
listed in the first row yields a range of virial central black hole
mass estimate of $1.8 \times 10^{9}M_{\odot} < M_{bh} < 4.9 \times
10^{9}M_{\odot}$. The quasar was observed on 8/15/2014 with the
G160M grating and on 8/20/2014 with the G130M. grating. These data
were down loaded from MAST (Mikulski Archive for Space Telescopes).
These spectra have never been discussed in the literature
previously.
\par In the top frame of Figure 1 is the EUV spectrum corrected for
Galactic extinction using the method of \citet{car89} with $R_{V}
=3.1$ and the visual extinction from the NASA Extragalactic Database
(NED). The plotted wavelength is the quasar rest frame wavelength,
$\lambda_{e}$, in order to facilitate the identification of the
emission lines. The flux density is that which is measured at earth,
designated as $F_{\lambda_{o}}$. From \citet{pun14}, the luminosity
near the peak of the spectral energy distribution at $\lambda_{e} =
1100$\AA\ (quasar rest frame wavelength), provides a robust
estimator of the bolometric luminosity associated with the thermal
accretion flow, $L_{\mathrm{bol}}$,
\begin{equation}
L_{bol} \approx 3.8 F_{\lambda_{e}}(\lambda_{e} = 1100 \AA)\;.
\end{equation}
Note that his estimator does not include reprocessed radiation in
the infrared from distant molecular clouds. This would be double
counting the thermal accretion emission that is reprocessed at
mid-latitudes \citep{dav11}. If the molecular clouds were not
present, this radiation would be directed away from the line of
sight to Earth. However it is reradiated back into the line of sight
towards Earth and this combines with the radiation that has a direct
line of sight to Earth from the thermal accretion flow. As such, it
would skew our estimate of the bolometric luminosity of the thermal
accretion flow and needs to be subtracted from the broadband
spectral energy distribution. Applying this equation to the spectrum
in Figure 1 indicates a very high luminosity, $L_{\mathrm{bol}}
\approx 1.3 \times 10^{47}\,\mathrm{ergs/s}$. However, as noted
above, $M_{bh}$ is very large, so the Eddington luminosity is large
$ 2.3 \times 10^{47}\,\mathrm{ergs/s} < L_{Edd} < 6.2 \times
10^{47}\,\mathrm{ergs/s}$. Even so, the data indicates a relatively
high Eddington ratio, $ 0.21 < L_{\mathrm{bol}}/L_{Edd} < 0.57 $.
\par The G130M raw spectrum was affected by a Lyman limit system
just above the Galactic Ly$\alpha$ region of contamination, making
it difficult to detect. The absorption begins at $\lambda_{o}
\approx 1240 \AA$ (Todd Tripp private communication 2016). A
correction was made for the depressed emission using a $\nu^{-3}$
opacity law. The smoothest extension of the continuum to short
wavelengths is attained with a maximum flux depression of 30\% to
35\%. A correction of 35\% was used to produce Figure 1. The
resulting continuum fit was made with $\alpha_{EUV} = 2.15$ and it
is plotted with the red dashed line.
\begin{table*}
\caption{Central Black Hole Mass
Estimate}\small{\begin{tabular}{cccccccc} \hline
 Source &  $\log{\lambda L_{\lambda}(\lambda =5100 \AA\,)}$ & FWHM H$\beta$ & $\log{\lambda L_{\lambda}(\lambda =3000 \AA\,)}$ & FWHM MgII & $\log{M_{bh}/M_{\odot}}$ & $\log{M_{bh}/M_{\odot}}$ \\
 & (ergs/s)   & (km/s) & (ergs/s)& (km/s)& H$\beta$ & Mg II\\
\hline
 3C95 & $45.04 \pm 0.11^{1}$ & $9030 \pm 720^{1}$ & $46.12 \pm 0.09^{2}$ & $5841 \pm 1168^{2}$ &$9.35 \pm 0.09^{3}$ & $9.53 \pm 0.16^{4}$\\

PKS0405-123 & $45.60 \pm 0.11^{1}$ & $4720 \pm 450^{1}$ & $46.40 \pm 0.09^{2}$ & $3933 \pm 787^{2}$ &$9.05 \pm 0.10^{3}$ & $9.31 \pm 0.16^{4}$\\

 3C57 & $45.17 \pm 0.09^{5}$ & $4500 \pm 470^{5}$ & $45.88 \pm 0.09^{5}$ & $3056 \pm 245^{5}$ &$8.80 \pm 0.10^{3}$ & $8.85 \pm 0.08^{4}$\\

 \hline

\end{tabular}}
\footnotesize{References: 1.\citet{mar03}  2.\citet{dec11}
3.\citet{ves06}  4.\citet{she12} 5.\citet{sul15}  }
\end{table*}
\subsection{PKS 0405-123} PKS 0405-123 is a classical triple radio
source at a redshift of z = 0.573. It is the most luminous EUV
source in this study. It has two strong radio lobes and a bright
flat spectrum core. The source is rather compact only 120 kpc from
end to end \citep{hut96}. Thus, the line of sight to the nucleus is
likely somewhat more polar than for 3C 95. The FWHM of the broad
component of H$\beta$ is 4720 km/s which might be moderated due to a
more polar line of sight than in 3C 95. Table 1 indicates a range of
virial central black hole mass estimates of $ 9.6 \times
10^{8}M_{\odot} < M_{bh} < 3.0 \times 10^{9}M_{\odot}$. The mass
might be slightly underestimated due to the effects of a more polar
line of sight.
\par In order to eliminate possible artifacts from continuum variability, it is desirable to
obtain simultaneous G130M and G160M observations. This was
accomplished by the observation on 12/21/2009 and these data were
downloaded from MAST.
\par In the middle frame of Figure 1 the EUV spectrum corrected for
Galactic extinction using the method of \citet{car89} with $R_{V}
=3.1$ and the visual extinction from NED is plotted. Equation (1)
and the spectrum in Figure 1 indicates a very high luminosity,
$L_{\mathrm{bol}} \approx 3.1 \times 10^{47}\,\mathrm{ergs/s}$. From
the black hole mass estimates, the Eddington ratio is extremely
high, $ 0.83 < L_{\mathrm{bol}}/L_{Edd} < 2.56 $. Even considering
possible more polar line of sight effects in the black hole mass
estimate, one concludes that the Eddington ratio is $\sim100\%$.
This seems very extreme, so one must also ask if there is a
significant contribution to the continuum from the optical high
frequency tail of the synchrotron emission associated with the radio
core? There are two strong pieces of evidence against this
possibility. First of all, the optical polarization is very low,
0.5\%, which is atypical of blazar dilution of the optical quasar
spectrum in low redshift sources \citep{wil92}. Secondly, the EUV
spectrum in Figure 1 shows very strong emission lines for which the
equivalent width does not appear to be reduced by a synchrotron
background component. Super-Eddington accretion rates were
previously found by the completely independent method of fitting
accretion disk models to the broadband spectrum \citep{mal83}. PKS
0405-123, with its powerful radio lobes (see the next section), is
perhaps the most extreme counter-example to the notion that radio
loud quasars are associated with low Eddington rates \citep{bor02}.
This is apparently a very remarkable radio source worthy of further
detailed study. The EUV continuum fit is indicated by the dashed red
line and it is estimated that $\alpha_{EUV} = 1.80$.
\subsection{3C 57} 3C 57 is a luminous quasar at z = 0.671 with a
complex radio morphology. It has a double steep spectrum nucleus
(two bright components separated by $\approx 9$ kpc, located within
the host galaxy) and a low luminosity radio lobe
\citep{rei99,hut88}. This cryptic radio structure will be analyzed
in detail in the next section. The results in Table 1 indicate a
range of virial central black hole mass estimates, $7.0 \times
10^{8}M_{\odot} < M_{bh} < 9.3 \times 10^{8}M_{\odot}$. The quasar
was observed on 8/17/2011 with the G160M grating and on 8/19/2011
with the G130M. grating. These data were down loaded from MAST.
\par In the bottom frame of Figure 1, the EUV spectrum corrected for
Galactic extinction using the method of \citet{car89} with $R_{V}
=3.1$ and the visual extinction from NED is plotted. Equation (1)
and the spectrum in Figure 1 indicates a high luminosity,
$L_{\mathrm{bol}} \approx 8.2 \times 10^{46}\,\mathrm{ergs/s}$. From
the black hole mass estimate, the Eddington ratio is extremely high,
$ L_{\mathrm{bol}}/L_{Edd} \approx 0.81 \pm 0.11 $. The EUV
continuum fit is indicated by the dashed red line and it is
estimated that $\alpha_{EUV} = 1.65$. An attempt was made to
consider the possibility that the decline in the flux density at
wavelengths just above the excised Galactic Ly$\alpha$ region is a
consequence of a Lyman limit system (at approximately with
$\lambda_{o} \approx 1235 \AA$) as was done for 3C 95. In this
scenario, even a 15 - 20\% decrement correction degraded the
continuum extension significantly from just above the excised
Galactic Ly$\alpha$ region of the spectrum to that below the excised
Galactic Ly$\alpha$ region of the spectrum. Any Lyman limit system
correction will raise the continuum at wavelengths just above the
excised Galactic Ly$\alpha$ region of the spectrum more than it will
below the excised region. This exaggerates the already existing
small dip in the continuum in this region (see Figure 1), making the
continuum look very unnatural. It is concluded that the decline in
the flux density at the short wavelength end of the spectrum
relative to the flux density at wavelengths just above the excised
Galactic Ly$\alpha$ region represents the expected drop in the
emission line flux short-ward of the blue wing of the very broad
blend of NIV, NeVIII and OIV that is generally very strong in all of
the spectra in Figure 1.
\begin{figure}
\includegraphics[width= 80 mm, angle=0]{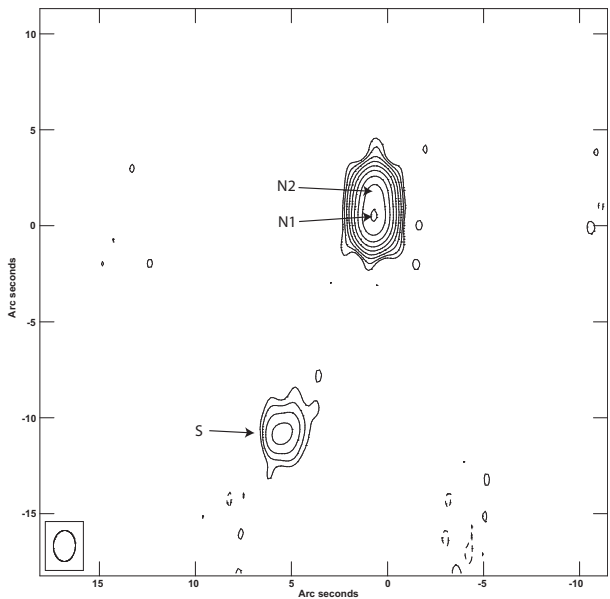}
\caption{A VLA radio image of 3C 57 at 1.5 GHz. Note the lobe
emission to the southeast and the partially resolved northern
component. The northern component is fit by two elliptical Gaussian
models, denoted by N1 and N2, separated by 1.35". The contours start
at 6.6 mJy/beam and increase in factors of 2 to a maximum of 1690
mJy/beam. }
\end{figure}

\section{Estimating the Long Term Time Averaged Jet Power}
A method that allows one to convert 151 MHz flux densities,
$F_{151}$ (measured in Jy), into estimates of long term time
averaged jet power, $\overline{Q}$, (measured in ergs/s) is captured
by the formula derived in \citet{wil99,pun05}:
\begin{eqnarray}
 && \overline{Q} \approx [(\mathrm{\textbf{f}}/15)^{3/2}]1.1\times
10^{45}\left[X^{1+\alpha}Z^{2}F_{151}\right]^{0.857}\mathrm{ergs/s}\;,\\
&& Z \equiv 3.31-(3.65)\times\nonumber \\
&&\left[X^{4}-0.203X^{3}+0.749X^{2} +0.444X+0.205\right]^{-0.125}\;,
\end{eqnarray}
where $X\equiv 1+z$, $F_{151}$ is the total optically thin flux
density from the lobes (i.e., \textbf{contributions from Doppler
boosted jets or radio cores are removed}). This calculation of the
jet kinetic luminosity incorporates deviations from the overly
simplified minimum energy estimates into a multiplicative factor
\textbf{f} that represents the small departures from minimum energy,
geometric effects, filling factors, protonic contributions and low
frequency cutoff \cite{wil99}. The quantity, \textbf{f}, was further
determined to most likely be in the range of 10 to 20 \cite{blu00}.
In this paper we adopt the following cosmological parameters:
$H_{0}$=70 km/s/Mpc, $\Omega_{\Lambda}=0.7$ and $\Omega_{m}=0.3$.
Define the radio spectral index, $\alpha$, as
$F_{\nu}\propto\nu^{-\alpha}$. The formula is most accurate for
large classical double radio sources, thus we do not consider
sources with a linear size of less than 20 kpc which are constrained
by the ambient pressure of the host galaxy. Alternatively, one can
also use the independently derived isotropic estimator in which the
lobe energy is primarily inertial in form \citet{pun05}
\begin{eqnarray}
&&\overline{Q}\approx
5.7\times10^{44}(1+z)^{1+\alpha}Z^{2}F_{151}\,\mathrm{ergs/sec}\;.
\end{eqnarray}
Due to Doppler boosting on kpc scales, core dominated sources with a
very bright one sided jet (such as 3C 279 and most blazars) must be
treated with care \citep{pun95}. The best estimate is to take the
lobe flux density on the counter-jet side and multiply this value by
2 (bilateral symmetry assumption) and use this estimate for the flux
density in Equations (2) and (4).
\par For strong radio sources, the value of Equation (4) is
typically slightly less than that found in Equation (3) with
$\mathrm{\textbf{f}} = 10$ \citep{pun05}. Thus, we take Equation (4)
as a lower bound on the estimate of $\overline{Q}$ \citep{blu00}.
Likewise, Equation (2) with $\mathrm{\textbf{f}} = 20$ is the
maximum upper bound on the estimate of $\overline{Q}$ in the
following \citep{wil99,blu00}. The values of $\overline{Q}$ that we
list below are the average of the upper and lower bound. The
assigned uncertainty is the difference between the maximum bound and
the average.
\par 3C 95 is a large lobe dominated quasar, so the 24.8 Jy of flux at 160 MHz
found in NED is due almost entirely from the radio lobes. A typical
lobe spectral index of $\alpha \approx 0.8$ and Equations (2) - (4)
indicate that $\overline{Q} = 1.6 \pm 0.5\times 10^{46}$. PKS
0405-123 has a strong flat spectrum radio core. The multi-frequency
component decomposition from \citet{hut96} is used to subtract off
the core flux density as a function of wavelength. The lobe flux
density is 1580 mJy at 1.41 GHz. Using $\alpha \approx 0.8$ and
Equations (2) - (4) indicates that $\overline{Q} = 8.1 \pm 3.0\times
10^{45}$ erg/s.

\begin{table*}
\caption{Summary of VLA Observations of the Components of 3C57}
\begin{tabular}{cccccc}
\hline
 Component &   1.52 GHz Flux & 4.86 GHz Flux & 14.94 GHz Flux & $\alpha $ \\
     &  Density (mJy) & Density (mJy)/ & Density (mJy) &  \\
     &  1.6" beamwidth & 1.8" beamwidth & 1.9" beamwidth &  \\
     & RMS = 0.8 mJy/beam & RMS = 0.4 mJy/beam & RMS = 0.8 mJy/beam &  \\
     &  10/31/1988 & 3/18/1989 & 1/12/1987 &  \\
\hline
 Component N2  & $1076 \pm 55$ & $381 \pm 11$ & $155 \pm 5$ & 0.87  \\
 Component N1 & $1596 \pm 80$ & $765 \pm 22$  & $476 \pm 15$   & 0.54 \\
 Southern Lobe, S  & $106 \pm 6 $ & $ 29 \pm 2 $ & $9 \pm 1 $ &  1.07 \\

\end{tabular}
\end{table*}
\begin{figure*}
\includegraphics[width= 85 mm, angle= 0]{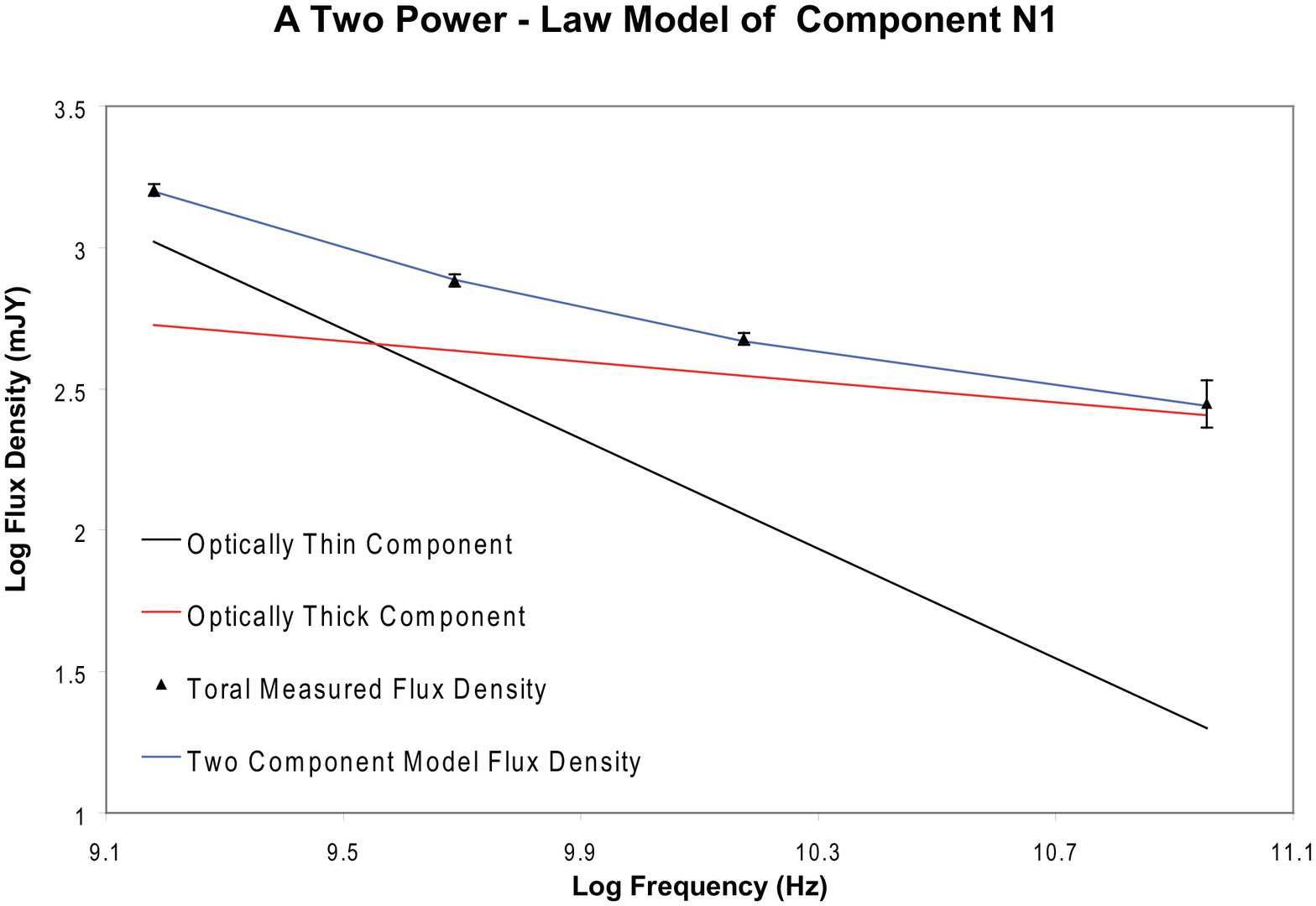}
\includegraphics[width= 85 mm, angle= 0]{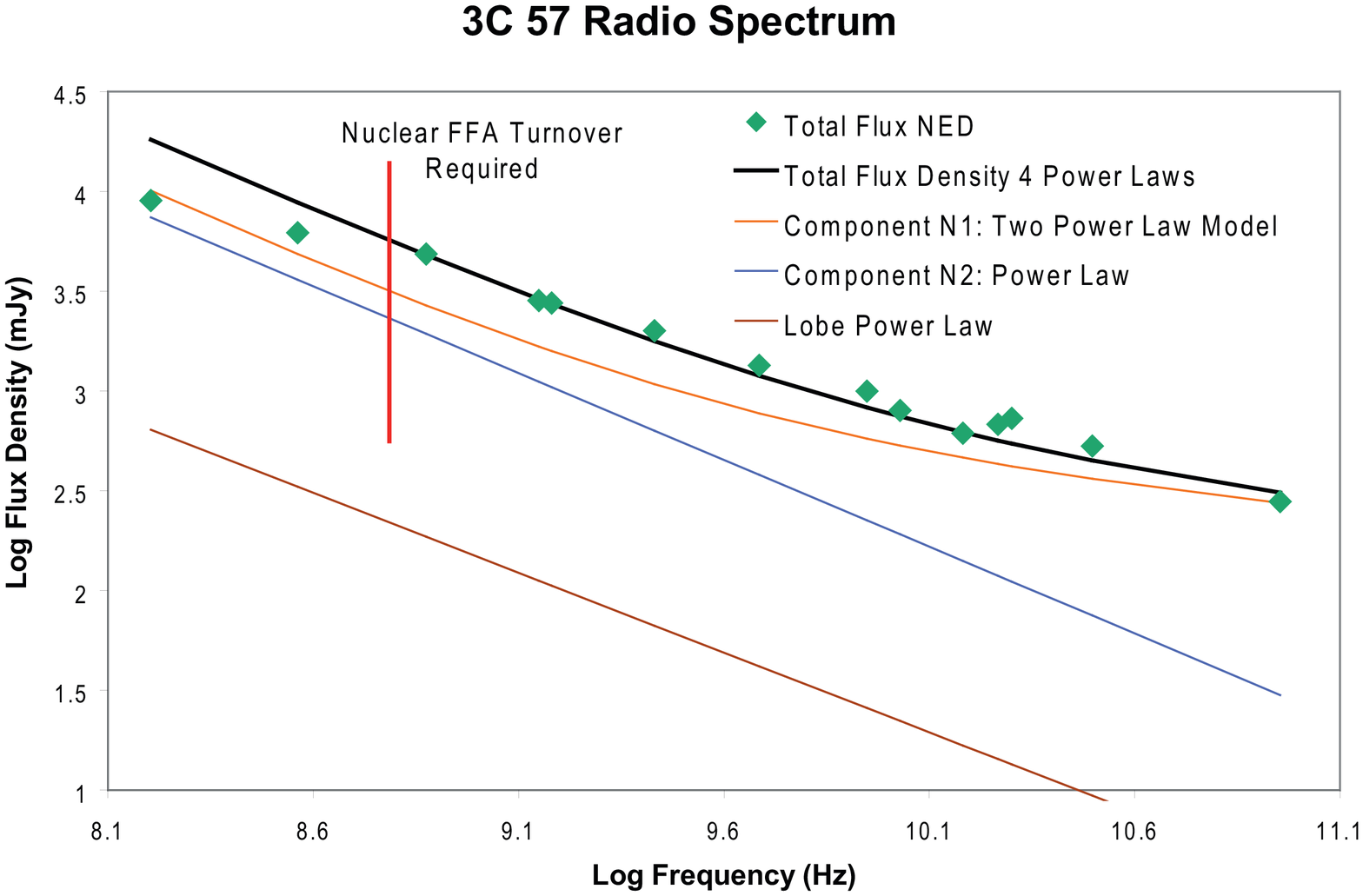}
\caption{The left frame shows that the two power-law component model
represents the four spectral data points of the component N1 from
Table 2, augmented by the 90 GHz flux density from Steppe et al.
1988. The right frame synthesizes the two power law component model
of component N1, the power law for component N2 in Table 2 and the
radio lobe with a spectral index $\alpha=0.8$ and 106 mJy at 1.52
GHz in order to facilitate a comparison to the observed data. This
is a four component model and the sum of the four components is the
dark black curve. The curve fits the data very well from 750 MHz to
90 GHz. Below 750 MHz there is clearly a spectral break required to
fit the low frequency spectrum. This is typically due to free-free
absorption (FFA).}
\end{figure*}
\par As mentioned earlier the radio morphology of 3C 57 is more
complicated. Figure 2 is a previously unpublished radio image of 3C
57 at 1.52 GHz from the VLA in A-Array from October 31, 1988. The
spatial resolution is reasonably well matched to published VLA
B-array at 4.86 GHz observations and the VLA C-array observations at
14.94 GHz \citep{rei99,bog94}. We have captured the the results of a
three component model of the source at the three frequencies in
Table 2. The components are a southern lobe and a a northern
component that we split as N1 and N2. The northern component is
partially resolved and is fit by two elliptical Gaussians separated
by 1.35". The columns (2) - (5) list the frequency of observation,
the beamwidth, the RMS noise level and the flux densities for each
component in the model. The matched resolution allows us to estimate
the spectral index of the components in a two component fit to the
northern component. The spectral indices are derived from the best
fit to the three data points for each component and are tabulated in
the last column. Component N1 is more luminous above 1 GHz and is
flatter spectrum. This is likely the true nucleus. The spectral
index for the lobe is not reliable. The flux densities at 4.86 GHz
and 14.94 GHz are not reliable for a diffuse lobe, since the surface
brightness is very low for this steep spectrum feature. Typically,
flux is missed in higher frequency observations of diffuse low
surface brightness structures. Thus, we still cannot rule out a more
typical spectral index of $\alpha = 0.8$ for the lobe emission.
\par The first thing that we analyze is component N1. We have three reliable flux density measurements. We note that
this component dominates the spectrum at 14.94 GHz and the spectrum
is the flattest of the three components. Therefore, it should
comprise the preponderance of flux at 90 GHz. The flux density at 90
GHz is $280 \pm 60$ mJy \citep{ste88}. In addition to the three
values in Table 2, 90 GHz provides a fourth frequency for which the
flux density is captured. The data in Table 1 already indicates a
spectral break at 4.86 GHz. So, we fit these four points with a
model comprised of two power-laws. There will be two amplitudes and
two power law indices representing 4 equations (the total flux at
each of the 4 frequencies) and four unknowns (the four parameters of
the two components model), so there exists a solution if the model
is a good approximation to the data. Based on many other steep
spectrum radio cores, one expects that a flat spectrum core might be
buried inside this feature. The left hand frame of Figure 3 is our
two component fit. We describe the optically thin (steep) component
of N1 by a flux density at 4.86 GHz of 339 mJy and a power law
spectral index of 0.97. We describe the optically thick (flat)
component of N1 by a flux density at 4.86 GHz of 431 mJy and a power
law spectral index of 0.18. This analysis clearly identifies N1 with
the central engine and the true nucleus.
\par An identification and characterization of the nucleus
elucidates the nature of the triple radio source in Figure 2. The
right frame of Figure 3 compares a four power law component model to
the radio data from NED augmented by the 90 GHz data and we correct
the 31.4 GHz data from NED, it should be $530 \pm 120$ mJy
\citep{gel81}. The four component model is the two power law
component model of N1 (from the left hand frame) and the power law
for N2 in Table 2 and the radio lobe with a spectral index
$\alpha=0.8$ and 106 mJy at 1.52 GHz. The sum of the four components
is the dark black curve. The curve fits the data very well from 750
MHz to 90 GHz. Below 750 MHz, the flux density of the model starts
to exceed the actual data, significantly. Below 750 MHz, there is
clearly a spectral break required to fit the low frequency spectrum.
This is likely a consequence free-free absorption (FFA), a
phenomenon commonly attributed to the spectral turnover of gigahertz
peaked radio sources \citep{ode98}. We also note a small discrepancy
in the fit near 20 GHz. This might be evidence of blazar-like
variability of the flat spectrum radio core.
\par One can compute the total radio luminosity of the compact steep
spectrum regions, if one assumes a strong spectral break between 365
MHz and 750 MHz based on the right hand frame of Figure 3. Secondly,
we assume that spectral ageing creates an enhanced downward
curvature in the spectrum at $\sim 50$ GHz. The radio luminosity,
$L_{\mathrm{radio}}$ is $\approx 2.3 \pm 0.2 \times 10^{44}$ erg/s
for the SNSC and $\approx 2.6 \pm 0.2 \times 10^{44}$ erg/s for the
northern component of the compact double.
\par The radio lobe $\approx$ 65 kpc to the southeast shows faint
extensions and an elongation back towards the nuclear region. This
is indicative of a hot spot at the termination of a radio jet.
Higher sensitivity 1.4 GHz B-Array observations with the VLA might
reveal some extended jet-like emission in this region. Comparing the
flux density of the faint radio lobe from 1.5 GHz, 5 GHz and 15 GHz
indicates a very steep spectrum with $\alpha > 1$. As noted above,
this result is likely skewed by faint diffuse emission being
resolved out at high frequency. However, a very steep spectrum is
not ruled out. In order to capture this possibility we extrapolate
the 1.52 GHz flux density of 106 mJy to 151 MHz (required for
Equations (2) and (4)) two different ways. For the upper bound from
Equation (2), we choose $\alpha=1$ and for the lower bound in
Equation (4), we use a more conventional $\alpha=0.8$. A 300 MHz VLA
A-array observation would resolve this ambiguity. However, as noted
in the discussion below Equation (4), we need to multiply our
calculation of $\overline{Q}$ by 2 in order to account for the lack
of an estimate for the northern radio jet. The northern radio jet
(component N2) is apparently thwarted by galactic gas and dissipates
$\approx 9$ kpc from the nucleus. Thus we estimate $\overline{Q} =
1.4 \pm 0.7 \times 10^{45} $ erg/s. In this interpretation, and
noting the value of $L_{\mathrm{radio}}$, above, the northern jet
dissipates between 21\% and 76\% of its energy as radio emission on
sub-galactic scales in component N2.

\begin{figure*}
\includegraphics[width= 85 mm, angle= 0]{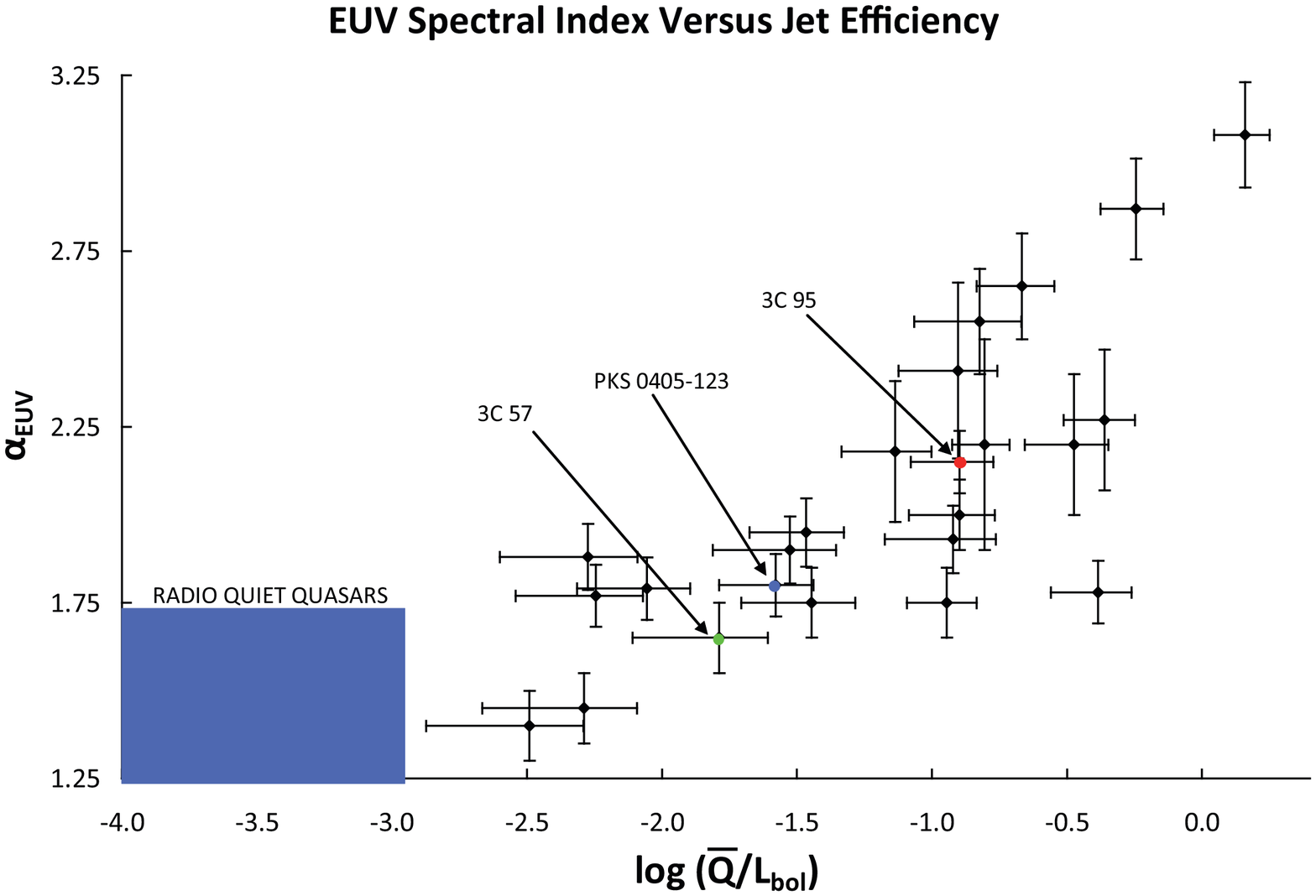}
\includegraphics[width= 85 mm, angle= 0]{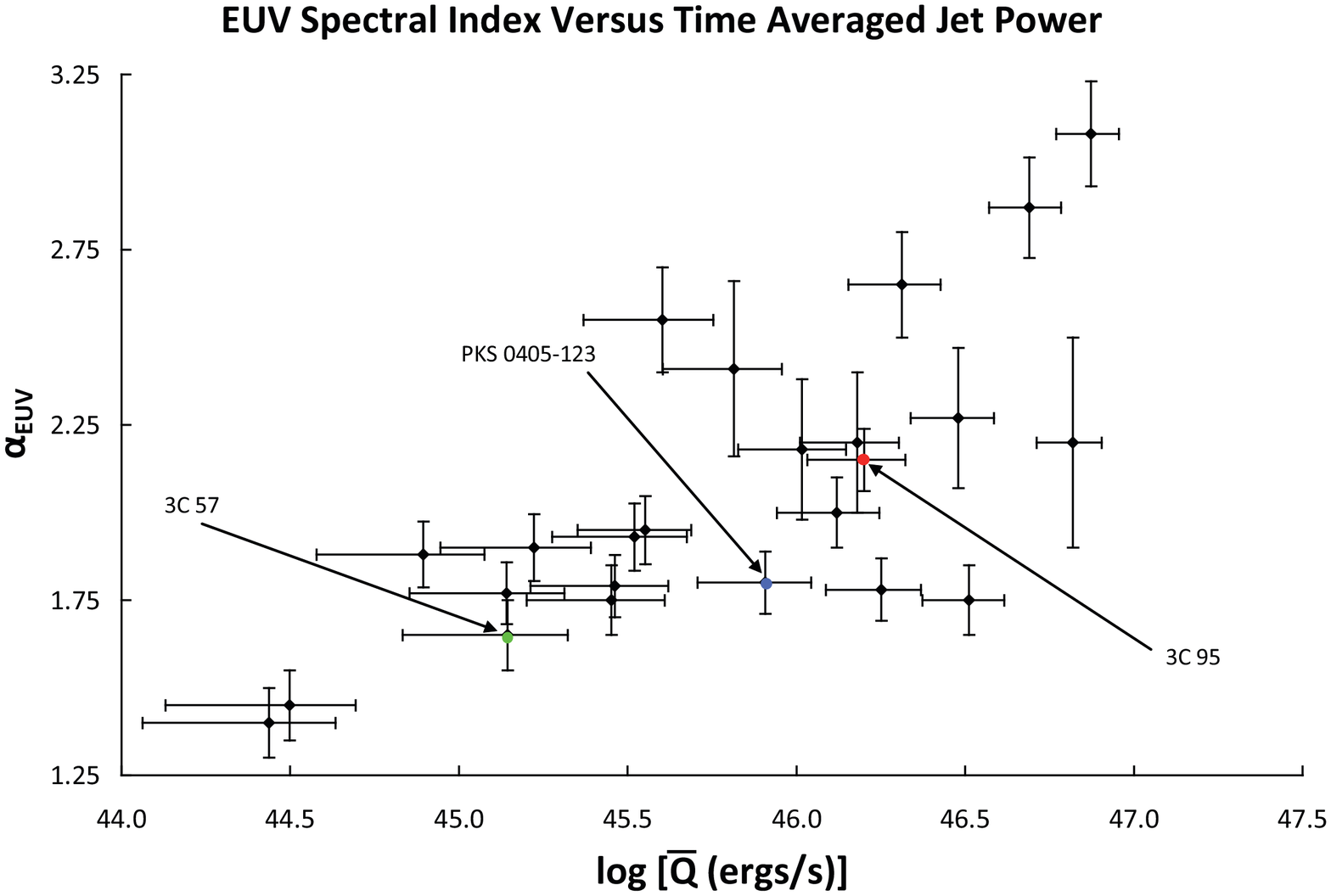}
\caption{A comparison of the scatter of the correlation of the EUV
spectral index, $\alpha_{EUV}$, with time averaged jet power,
$\overline{Q}$ (on the right), and the scatter of the correlation of
$\alpha_{EUV}$ with the time averaged jet power normalized by the
bolometric luminosity of the accretion flow
$\overline{Q}/L_{\mathrm{bol}}$ (on the left). These plots help to
visualize the partial correlation analysis described in the text.
Normalizing by the bolometric luminosity reduces the scatter at a
very high level of statistical significance.}
\end{figure*}
\par The large dissipation of jet power into radio luminosity is not
entirely unexpected. Recall the comment in the text following
Equation (3). Our estimators of jet power in Equations (2) and (4)
are technically accurate only for relaxed classical double radio
sources. These radio sources expand into a diffuse intra-galactic
medium Hence our restriction of applicability that the lobe must be
$> 20$ kpc from the central engine (outside of the host galaxy) in
order for us to apply this relationship to our radio sources.
Equation (2) yields a radio luminosity $\LA 1\%$ of the jet power. A
similar result has been found for in studies of classical relaxed
radio sources \citep{cav10,dal12}. Component N2 is $\approx 9$ kpc
from the central engine, and as we have noted previously, it is
likely located within the dense medium of the host galaxy. The work
required to displace dense nebular clouds and the intergalactic
medium, in general, likely involves MHD waves (including shock
waves) and instabilities that can be highly dissipative. This leads
to an enhanced radiative luminosity and Equations (2) and (3) will
always drastically overestimate the power of the jet if the
enveloping environment is dense. However, in the context of the
large radio luminosity of the northern component noted above, this
does not preclude the possibility that the central engine
underwent``a brief" (relative to the long lifetime of the radio
source) episode of elevated jet power $\approx 30,000$ years ago.
\section{Discussion}
This paper considers the EUV spectrum and radio properties of three
RLQs at the high end region of the quasar $L_{\mathrm{bol}}/L_{Edd}$
parameter space. In the left hand frame of Figure 4 is the scatter
plot of $\alpha_{EUV}$ - $\overline{Q}/L_{\mathrm{bol}}$ from
\citet{mar15} with our three new sources added. The three quasars
conform to the existing correlation. The exploration of the high end
of $L_{\mathrm{bol}}/L_{Edd}$ parameter space can be used to fortify
the statistics of the partial correlation analysis of \citet{pun15}
amongst the variables, $\overline{Q}/L_{\mathrm{bol}}$,
$\overline{Q}$ and $\alpha_{EUV}$ that indicated that the
correlation of $\overline{Q}/L_{\mathrm{bol}}$ and $\alpha_{EUV}$ is
fundamental and the correlation of $\overline{Q}$ and $\alpha_{EUV}$
is spurious. In particular, consider the Spearman partial
correlation of $\overline{Q}/L_{\mathrm{bol}}$ with
$\alpha_{\mathrm{EUV}}$ when $\overline{Q}$ is held fixed. The
partial correlation coefficient with the expanded sample is 0.594
(was 0.492 in \citet{pun15}) which correspondence to a statistical
significance of 0.998 (was 0.984). Conversely, the partial
correlation of $\overline{Q}$ with $\alpha_{\mathrm{EUV}}$ when
$\overline{Q}/L_{\mathrm{bol}}$ is held fixed has a a statistical
significance of only 0.188 (was 0.581). Thus, the addition of
quasars from the high end of $L_{\mathrm{bol}}/L_{Edd}$ parameter
space to the sample has accentuated the fact that the correlation of
$\overline{Q}/L_{\mathrm{bol}}$ and $\alpha_{EUV}$ is fundamental
and the correlation of $\overline{Q}$ and $\alpha_{EUV}$ is spurious
at a very high statistical significance level.
\par This is a very
important result at a fundamental physical level. For example, it
highlights the fact that the mass accretion rate is strongly coupled
to jet power in quasars that can support a relativistic jet. In
particular, the accretion rate regulates jet power in RLQs. As
discussed in detail in \citet{pun15}, this is a basic prediction of
magnetically arrested accretion scenarios for jet production in
quasars. This idea is predicated on the fact that large scale
magnetic flux is trapped within the inner accretion flow by ram
pressure. The rotating magnetic flux is the source of the
relativistic jet. More trapped magnetic flux means a stronger jet.
The trapped poloidal magnetic flux is vertical as it penetrates the
inner accretion flow (perpendicular to the plane on the inflow) in
some numerical simulations and models of magnetically arrested
accretion \citep{igu08,pun09}. Thus, as the large scale magnetic
flux forms the base of the radio jet, the thermal gas that is
displaced by the magnetic flux results in less EUV emission from the
inner accretion flow. Thus, a consistent explanation of the
$\overline{Q}/L_{\mathrm{bol}}$ and $\alpha_{EUV}$ correlation is
achieved given that there is significant vertical flux that threads
the inner accretion flow..

\end{document}